# Output-Constrained Controller with Fuzzy-Tuned Parameters for Overhead Cranes

Dawei Zhao[1], Kai Wang[1], Xianglong Zhou*[2], Xin Ma*[1], Lei Jia[1]

1. School of Control Science and Engineering, Shandong University, Jinan 250061, China

E-mail: maxin@sdu.edu.cn

2. Shandong Inspur Science Research Institute Co., Ltd., Jinan 250101, China

**Abstract:** This study proposes a fuzzy-adjusted nonlinear control method based on torque jitter output limit constraints for overhead crane systems with double pendulum effects. The proposed control method can effectively suppress swing and achieve precise positioning. Firstly, by enhancing the coupling relationship between the trolley displacement and swing angle, a composite signal with an error term was designed. Then, an energy-based Lyapunov function was constructed using the composite error signal, which incorporated a new formulation of the inertia matrix and potential energy function. Subsequently, using the backstepping method in conjunction with the hyperbolic tangent function, a controller with partial performance constraints was designed. In addition, to further enhance the system's dynamic performance, a fuzzy control scheme with online adjustable system parameters was designed. Finally, the stability of the system is proven using Lyapunov theory combined with LaSalle's invariance principle. Simulation results demonstrate that the proposed controller exhibits superior performance and robustness.

## 1 Introduction

Overhead cranes are commonly employed for cargo transportation in factories and ports [1]. As a typical underactuated mechanical system, overhead cranes have fewer independent control inputs than their degrees of freedom [2]. In recent years, the control of underactuated cranes has garnered significant interest from the robotics and control communities [3][4].

The demand for the development of automated driving technology for overhead cranes has been growing [5]. Crane systems frequently operate in demanding environments, such as harsh weather conditions and high-altitude work, which inevitably give rise to various challenges [6]. The primary goal in managing overhead crane systems is to achieve accurate positioning of both the trolley and the payload while minimizing oscillations that occur during movement. These oscillations are intrinsically linked to the underactuated nature of the crane system [7]. The control methods for overhead cranes are mainly divided into open-loop control and closed-loop control. Open-loop methods, such as the commonly used input shaping technique, are effective in eliminating payload oscillations and are straightforward to design [8]. Open-loop control methods do not rely on feedback, resulting in a simple system structure, but the control performance is not optimal. Currently, closed-loop control methods are more commonly used. To suppress payload swing, several energy-based approaches have been developed for overhead crane systems [9]. For example, Handling uncertainty is a major challenge in crane systems. Factors such as friction, external disturbances, and rope elasticity can complicate the control problem. Therefore, adaptive control is utilized to address these uncertainties [10][11]. Additionally, Model predictive control (MPC) is an advanced technique that has been applied to crane control [12]. In control theory, numerous studies have investigated the relationship between RK coefficients and the stability region associated with these methods, as well as the optimal selection of step sizes for RK approaches [13]. Additionally, in crane control, the PID control method is also commonly used [14]. In addition, there is also the energy passivity control method, which is based on the principles of energy conservation and dissipation [15]. However, most of the energy-based approach only considers overhead cranes with a single-pendulum effect, without accounting for the double-pendulum effect of the crane [16].

Due to the underactuated characteristics of the overhead crane, the trolley's torque can only directly control the crane's displacement, making it difficult to suppress the payload's swing angle. Displacement and swing angle are independent. To address this issue, this paper defines a coupling signal that combines displacement and swing angle. A vector composed of this signal replaces the generalized coordinate vector in the original dynamic model, resulting in a new form of the dynamic model. An energy function is then constructed, and by ensuring the derivative of the energy function is non-positive, the system's control law is derived. Given that the controller expression involves parameters requiring manual adjustment, a fuzzy logic adaptive strategy is designed to adjust the tunable parameters in the controller, thereby optimizing its performance. The system's stability is proven using Lyapunov theory and LaSalle's invariance principle.

*This work is supported by Marine Science and Technology Achievement Transfer and Transformation Center (Innovation and Entrepreneurship Community) "Unveiling the List and Leading the Way" Project of Shandong Province under Grant 2024JBGS05, Central Guidance for Local Scientific and Technological Development Funding Project of Shandong Province under Grant YDZX2023042, Key Research and Development Project of Shandong Province under Grant 2021CXGC010701.

Comparative experiments show that the controller demonstrates excellent control performance.

## 2 Dynamic model of overhead crane

The crane system exhibits a double pendulum effect under the following conditions: 1) when the hook mass is comparable to the load mass and cannot be neglected, and 2) when the load possesses substantial dimensions with non-uniform mass distribution, rendering it inapplicable to be idealized as a point mass. Therefore, the investigation of positioning and anti-sway control strategies for double-pendulum bridge cranes is of paramount research significance in advancing the hoisting industry.

The model of 3-DOF overhead crane system is shown in Fig.1. The parameters and variables are presented in Table 1.

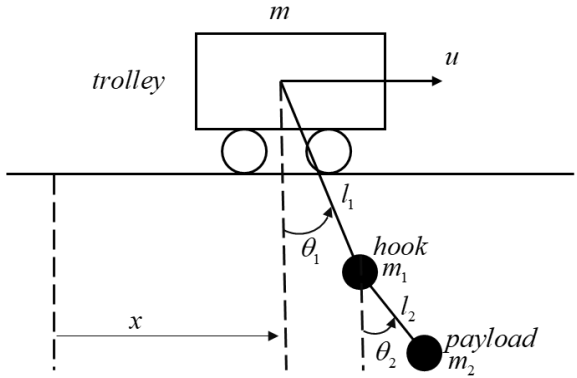

Fig. 1: 3-DOF overhead crane system model

Table 1: Parameters and Variables

| Para. | Parameter definition | Units |
|---|---|---|
| $m$ | Trolley mass | kg |
| $m_1$ | Hook mass | kg |
| $m_2$ | Payload mass | kg |
| $l_1$ | Length of the rope between the trolley and the hook | m |
| $l_2$ | Length of the rope between the hook and the payload | m |
| $x$ | Trolley displacement | m |
| $\theta_1$ | Hook swing angle | rad |
| $\theta_2$ | Payload swing angle | rad |

Using the Lagrange equation to model the dynamics of an overhead crane with double-pendulum effects, the following dynamic equations are obtained:

$$(m+m_1+m_2)\ddot{x}+(m_1+m_2)l_1\cos\theta_1\ddot{\theta}_1+m_2l_2\cos\theta_2\ddot{\theta}_2 \\ -(m_1+m_2)l_1\sin\theta_1\dot{\theta}_1^2-m_2l_2\sin\theta_2\dot{\theta}_2^2=u \quad (1)$$

$$(m_1+m_2)l_1\cos\theta_1\ddot{x}+(m_1+m_2)l_1^2\ddot{\theta}_1+m_2l_1l_2\cos(\theta_1-\theta_2)\ddot{\theta}_2 \\ +m_2l_1l_2\sin(\theta_1-\theta_2)\dot{\theta}_2^2+(m_1+m_2)gl_1\sin\theta_1=0 \quad (2)$$

$$m_2l_2\cos\theta_2\ddot{x}+m_2l_1l_2\cos(\theta_1-\theta_2)\ddot{\theta}_1+m_2l_2^2\ddot{\theta}_2 \\ -m_2l_1l_2\sin(\theta_1-\theta_2)\dot{\theta}_1^2+m_2gl_2\sin\theta_2=0. \quad (3)$$

Considering the complexity of the 3-DOF overhead crane system, and the fact that the swing angle is very small near the equilibrium point, the following approximation is used:

$$\cos\theta_i\approx 1, \cos(\theta_1-\theta_2)\approx 1, \sin(\theta_1-\theta_i)\dot{\theta}_i\approx 0, \\ \sin\theta_i\approx\theta_i, \quad i=1,2. \quad (4)$$

The simplified dynamic model is converted into a compact matrix form as follows:

$$M(q)\ddot{q}+C(q,\dot{q})\dot{q}+G(q)=U \quad (5)$$

where $M(q)$ is the symmetric inertia matrix. $C(q,\dot{q})$ denote the Centrifugal-Coriolis matrix. $G(q)$ describes the gravity vector. $q$ is the generalized coordinate vector. $U$ is the input vector. These three matrices are shown as

$$M(q)=\begin{bmatrix} m+m_1+m_2 & (m_1+m_2)l_1 & m_2l_2 \\ (m_1+m_2)l_1 & (m_1+m_2)l_1^2 & m_2l_1l_2 \\ m_2l_2 & m_2l_1l_2 & m_2l_2^2 \end{bmatrix},$$

$$C(q,\dot{q})=\begin{bmatrix} 0 & -(m_1+m_2)l_1\theta_1\dot{\theta}_1 & -m_2l_2\theta_2\dot{\theta}_2 \\ 0 & 0 & 0 \\ 0 & 0 & 0 \end{bmatrix},$$

$$G(q)=\begin{bmatrix} 0 \\ (m_1+m_2)gl_1\theta_1 \\ m_2gl_2\theta_2 \end{bmatrix}, \quad q=[x\ \theta_1\ \theta_2]^T \text{ and } U=[u\ 0\ 0]^T.$$

Considering that during the operation of the bridge crane, the payload is always positioned directly beneath the trolley, the following reasonable assumptions are made:

$$-\frac{\pi}{2}<\theta_i<\frac{\pi}{2}(i=1,2), \forall t\geq 0. \quad (6)$$

## 3 Construction of energy function

To enhance the coupling between the trolley displacement and the payload swing angle in the two-degree-of-freedom overhead crane, this paper constructs a novel energy function by combining the payload swing angle and the trolley position, and defines the following composite signal:

$$\varepsilon=x-K_l\,l_1(\int_0^t\sin\theta_1 d\tau+\int_0^t\sin\theta_2 d\tau) \quad (7)$$

where $K_l$ is a positive gain parameter, and $l_1$ is the rope length between the trolley and the hook.

Then, define $e_\varepsilon$ as the error:

$$e_\varepsilon=\varepsilon-x_d=x-x_d-K_l\,l_1(\int_0^t\sin\theta_1 d\tau+\int_0^t\sin\theta_2 d\tau). \quad (8)$$

To conveniently construct a new form of the dynamic model, the error vector and its derivative are defined as follows:

$$e=[e_\varepsilon\ \theta_1\ \theta_2]^T \\ =[x-x_d-K_l\,l_1(\int_0^t\sin\theta_1 d\tau+\int_0^t\sin\theta_2 d\tau)\ \theta_1\ \theta_2]^T \quad (9)$$

$$\dot{e}=[\dot{e}_\varepsilon\ \dot{\theta}_1\ \dot{\theta}_2]^T=[\dot{x}-K_l l_1(\sin\theta_1+\sin\theta_2)\ \dot{\theta}_1\ \dot{\theta}_2]^T. \quad (10)$$

Furthermore,

$$\ddot{e}=[\ddot{e}_\varepsilon\ \ddot{\theta}_1\ \ddot{\theta}_2]^T \\ =[\ddot{x}-K_l l_1(\cos\theta_1\dot{\theta}_1+\cos\theta_2\dot{\theta}_2)\ \ddot{\theta}_1\ \ddot{\theta}_2]^T. \quad (11)$$

By replacing the generalized coordinate vector in the original dynamic model with the error vector, the new form of the dynamic model is obtained as follows:

$$M(q)\ddot{e}+C(q,\dot{q})\dot{e}+G(q)+R(q,\dot{q})=U \quad (12)$$

where

$$R(q,\dot{q})=K_l\begin{bmatrix} (m+m_1+m_2)l_1(\cos\theta_1\dot{\theta}_1+\cos\theta_2\dot{\theta}_2) \\ (m_1+m_2)l_1^2(\cos\theta_1\dot{\theta}_1+\cos\theta_2\dot{\theta}_2) \\ m_2l_1l_2(\cos\theta_1\dot{\theta}_1+\cos\theta_2\dot{\theta}_2) \end{bmatrix}.$$

The energy function of a mechanical system is typically expressed as the sum of kinetic and potential energies. Therefore, consider the following positive definite energy function:

$$V_d(t) = \frac{1}{2}\dot{e}^T M_I \dot{e} + E_p(e) \qquad (13)$$

where $M_I$ is the positive definite inertia matrix to be designed, and $E_p(e)$ is the desired potential energy function.

Taking the derivative of the above Eq. (13) gives:

$$\dot{V}_d(t) = \dot{e}^T(M_I \ddot{e} + \frac{\partial E_p(e)}{\partial e}). \qquad (14)$$

To ensure system stability, we set the derivative of the energy function to be non-positive:

$$\dot{V}_d(t) = -\dot{e}^T D(q)\dot{e} \qquad (15)$$

where $D(q)$ is the desired damping matrix to be designed.

From Eq. (14) and (15), we can obtain:

$$M_I \ddot{e} + \frac{\partial E_p(e)}{\partial e} + D(q)\dot{e} = 0. \qquad (16)$$

Multiplying both sides of the above equation by the inverse matrix $M_I^{-1}$ on the left yields:

$$\ddot{e} = -M_I^{-1}\frac{\partial E_p(e)}{\partial e} - M_I^{-1}D(q)\dot{e}. \qquad (17)$$

Substitute Eq. (17) into Eq. (12), we can obtain:

$$U = C(q,\dot{q})\dot{e} + G(q) + R(q,\dot{q}) - M(q)M_I^{-1}\frac{\partial E_p(e)}{\partial e} \qquad (18)$$
$$- M(q)M_I^{-1}D(q)\dot{e}.$$

For Eq. (18), to facilitate the design of the desired inertia matrix $M_I$, the desired potential energy function $E_p(e)$ and the damping matrix $D(q)$, thus:

$$\alpha U = \alpha[C(q,\dot{q})\dot{e} + G(q) + R(q,\dot{q}) - M(q)M_I^{-1}\frac{\partial E_p(e)}{\partial e} \qquad (19)$$
$$- M(q)M_I^{-1}D(q)\dot{e}] = 0$$

where $\alpha = [0\ -l_2\ l_1]$ and Eq. (19) contains $M_I$, $E_p(e)$ and $D(q)$. To satisfy the above Eq. (19), let

$$\alpha[C(q,\dot{q})\dot{e} + R(q,\dot{q}) - M(q)M_I^{-1}D(q)\dot{e}] = 0 \qquad (20)$$

$$\alpha[G(q) - M(q)M_I^{-1}\frac{\partial E_p(e)}{\partial e}] = 0. \qquad (21)$$

To simplify the design process, the desired inertia matrix is set as the identity matrix:

$$M_I = M_I^{-1} = \begin{bmatrix} 1 & 0 & 0 \\ 0 & 1 & 0 \\ 0 & 0 & 1 \end{bmatrix}.$$

It is noted that

$$\alpha C(q,\dot{q})\dot{e} \equiv 0. \qquad (22)$$

Therefore, Eq. (20) and (21) can be simplified as:

$$\alpha[R(q,\dot{q}) - M(q)D(q)\dot{e}] = 0 \qquad (23)$$

$$\alpha[G(q) - M(q)\frac{\partial E_p(e)}{\partial e}] = 0. \qquad (24)$$

For Eq. (23), let $D(q) = D_1(q) + D_2(q)$.

Thus, Eq. (23) can be transformed into

$$\alpha[R(q,\dot{q}) - M(q)D_1(q)\dot{e} + M(q)D_2(q)\dot{e}] = 0. \qquad (25)$$

To satisfy Eq. (25), let:

$$\alpha[R(q,\dot{q}) - M(q)D_1(q)\dot{e}] = 0 \qquad (26)$$

$$\alpha M(q)D_2(q)\dot{e} = 0. \qquad (27)$$

Eq. (26) gives:

$$[0\ -l_2\ l_1]\{K_I \begin{bmatrix} (m+m_1+m_2)l_1(\cos\theta_1\dot{\theta}_1+\cos\theta_2\dot{\theta}_2) \\ (m_1+m_2)l_1^2(\cos\theta_1\dot{\theta}_1+\cos\theta_2\dot{\theta}_2) \\ m_2 l_1 l_2(\cos\theta_1\dot{\theta}_1+\cos\theta_2\dot{\theta}_2) \end{bmatrix} -$$
$$\begin{bmatrix} m+m_1+m_2 & (m_1+m_2)l_1 & m_2 l_2 \\ (m_1+m_2)l_1 & (m_1+m_2)l_1^2 & m_2 l_1 l_2 \\ m_2 l_2 & m_2 l_1 l_2 & m_2 l_2^2 \end{bmatrix} D_1(q) \begin{bmatrix} \dot{e}_\varepsilon \\ \dot{\theta}_1 \\ \dot{\theta}_2 \end{bmatrix}\} = 0. \qquad (28)$$

It is easy to obtain $D_1(q)$ that satisfies Eq. (28):

$$D_1(q) = K_I \begin{bmatrix} 0 & 0 & 0 \\ 0 & \cos\theta_1 & \cos\theta_2 \\ 0 & -\cos\theta_2 & \cos\theta_2 \end{bmatrix}. \qquad (29)$$

Based on Eq. (27), design $D_2(q)$ in the following form:

$$D_2(q) = K_d M^{-1}(q)\beta^T \beta M^{-1}(q)$$
$$= -\frac{K_d}{m^2 l_1^2}\begin{bmatrix} l_1^2 & -l_1 & 0 \\ -l_1 & 1 & 0 \\ 0 & 0 & 0 \end{bmatrix} \qquad (30)$$

where $\beta = [1\ 0\ 0]$ and $K_d$ is a positive gain.

Thus, $D(q)$ can be expressed as follows:

$$D(q) = K_I \begin{bmatrix} 0 & 0 & 0 \\ 0 & \cos\theta_1 & \cos\theta_2 \\ 0 & -\cos\theta_2 & \cos\theta_2 \end{bmatrix} - \frac{K_d}{m^2 l_1^2}\begin{bmatrix} l_1^2 & -l_1 & 0 \\ -l_1 & 1 & 0 \\ 0 & 0 & 0 \end{bmatrix}. \qquad (31)$$

Next, discuss the selection of $E_p(e)$.

Substitute $M(q)$ and $G(q)$ into Eq. (24):

$$[0\ -l_2\ l_1]\{\begin{bmatrix} 0 \\ (m_1+m_2)gl_1\theta_1 \\ m_2 g l_2 \theta_2 \end{bmatrix}$$
$$- \begin{bmatrix} m+m_1+m_2 & (m_1+m_2)l_1 & m_2 l_2 \\ (m_1+m_2)l_1 & (m_1+m_2)l_1^2 & m_2 l_1 l_2 \\ m_2 l_2 & m_2 l_1 l_2 & m_2 l_2^2 \end{bmatrix}\begin{bmatrix} \frac{\partial E_p(e)}{\partial e_\varepsilon} \\ \frac{\partial E_p(e)}{\partial \theta_1} \\ \frac{\partial E_p(e)}{\partial \theta_2} \end{bmatrix}\} = 0. \qquad (32)$$

From Eq. (32), we can obtain:

$$\frac{\partial E_p(e)}{\partial \theta_1} = (\frac{g}{l_1} + \frac{m_2 g}{m_1 l_1})\theta_1 - \frac{1}{l_1}\frac{\partial E_p(e)}{\partial e_\varepsilon} - \frac{m_2 g}{m_1 l_1}\theta_2. \qquad (33)$$

Let:

$$\frac{\partial E_p(e)}{\partial e_\varepsilon} = K_p \tanh(e_\varepsilon - \frac{\theta_1+\theta_2}{l_1}). \qquad (34)$$

From Eq. (33) and (34), we can obtain:

$$\frac{\partial E_p(e)}{\partial \theta_1} = (\frac{g}{l_1} + \frac{m_2 g}{m_1 l_1})\theta_1 - \frac{1}{l_1}K_p \tanh(e_\varepsilon - \frac{\theta_1+\theta_2}{l_1}) \qquad (35)$$
$$- \frac{m_2 g}{m_1 l_1}\theta_2.$$

Thus, the expression for $E_p(e)$ can be derived as follows:

$$E_p = \frac{1}{2}(\frac{g}{l_1} + \frac{m_2 g}{m_1 l_1})\theta_1^2 + K_p \ln\cosh(e_\varepsilon - \frac{\theta_1+\theta_2}{l_1}) \qquad (36)$$
$$+ \frac{3-\theta_1\theta_2}{m_1 l_1}m_2 g.$$

It can also be obtained that

$$\frac{\partial E_p(e)}{\partial \theta_2} = -\frac{1}{l_1} K_p \tanh(e_\varepsilon - \frac{\theta_1 + \theta_2}{l_1}) - \frac{m_2 g}{m_1 l_1} \theta_1. \quad (37)$$

## 4 Controller design

To obtain the expression of the control input, multiplying both sides of Eq. (18) by $\gamma = [1 \ 0 \ 0]$, we get

$$\gamma U = \gamma [C(q,\dot{q})\dot{e} + G(q) + R(q,\dot{q}) \\ - M(q)M_I^{-1} \frac{\partial E_p(e)}{\partial e} - M(q)M_I^{-1} D(q)\dot{e}]. \quad (38)$$

The expression of the control input can be solved as follows from Eq. (38):

$$u = -K_p \left(m - \frac{m_2 l_2}{l_1}\right) \tanh\left(\frac{e_\varepsilon - (\theta_1 + \theta_2)}{l_1}\right) - K_d \left(\frac{\dot{e}_\varepsilon}{m} - \frac{\dot{\theta}_1}{ml_1}\right) \\ + K_l \left[ml_1(\cos\theta_1 \dot{\theta}_1 + \cos\theta_2 \dot{\theta}_2) - m_2 l_2 \cos\theta_2 (\dot{\theta}_2 - \dot{\theta}_1)\right] \\ - \left(m_1 g + \frac{m_2^2 g}{m_1} - \frac{m_2^2 g l_2}{m_1 l_1} + 2m_2 g\right) \theta_1 + \left(\frac{m_2^2 g}{m_1} + m_2 g\right) \theta_2 \\ - (m_1 l_1 + m_2 l_1) \theta_1 \dot{\theta}_1^2 - m_2 l_2 \theta_2 \dot{\theta}_2^2. \quad (39)$$

When the trolley is at the initial position, $\dot{e}_\varepsilon = 0$, $\theta_1 = 0$, $\theta_2 = 0$, $\dot{\theta}_1 = 0$, $\dot{\theta}_2 = 0$, $e_\varepsilon = -x_d$. Therefore, $u(0) = -K_p \left(m - \frac{m_2 l_2}{l_1}\right) \tanh\left(\frac{-x_d}{l_1}\right)$ implies that the controller's control action is bounded in the initial state. Even if $x_d$ is large, the absolute value of the controller's output can still be controlled within $K_p \left(m - \frac{m_2 l_2}{l_1}\right)$. That is to say, a hyperbolic tangent function is incorporated to partially constrain the controller output, effectively mitigating abrupt torque transitions during trolley acceleration phases.

## 5 Fuzzy logic adaptive strategy design

From the controller expression, it is evident that the controller contains three parameters that need to be manually tuned, namely $K_p$, $K_d$ and $K_l$. To achieve auto-tuning of the controller parameters and improve the control performance and accuracy, this paper designs a fuzzy logic adaptive strategy to adjust these three parameters online.

The inputs to the fuzzy logic adaptive strategy are the displacement error $e$ and the derivative of the displacement error $\dot{e}$, while the outputs of the fuzzy logic adaptive strategy are the increments of the three parameters ($K_p$, $K_d$ and $K_l$). The fuzzy sets for $e$, $\dot{e}$, $K_p$, $K_d$, $K_l$ are both {NB, NM, NS, ZE, PS, PM, PB}, and the domains of the input and output variables of the controller are $e \in (-1,1)$, $\dot{e} \in (-0.5, 0.5)$, $\triangle K_p \in (-0.25, 0.25)$, $\triangle K_d \in (-10,10)$, $\triangle K_l \in (-0.05, 0.05)$. Since the output of the fuzzy logic adaptive strategy represents the increments of the controller parameters, we have:

$$K_p = K_{p0} + \triangle K_p \quad (40)$$
$$K_d = K_{d0} + \triangle K_d \quad (41)$$
$$K_l = K_{l0} + \triangle K_l \quad (42)$$

where $K_{p0}$, $K_{d0}$ and $K_{l0}$ represent the initial values of the three controller parameters, and $\triangle K_p$, $\triangle K_d$ and $\triangle K_l$ are the increments of the three controller parameters.

The fuzzy rule table is shown in Table 2.

Table 2: Fuzzy Rule Table

| $\triangle K_p / \triangle K_d / \triangle K_l$ | | $e_2$ | | | | | | |
|---|---|---|---|---|---|---|---|---|
| | | NB | NM | NS | ZE | PS | PM | PB |
| $\dot{e}_2$ | NB | PB/PS/NB | PB/PS/NB | PM/ZE/PB | PM/ZE/ZE | PS/ZE/PB | PS/PB/NB | ZE/PB/NB |
| | NM | PB/NS/NB | PB/NS/NB | PM/NS/PB | PM/NS/ZE | PS/ZE/PB | ZE/NS/NB | ZE/PM/NB |
| | NS | PM/NB/NB | PM/NB/NB | PM/NM/PB | PS/NS/ZE | ZE/ZE/PB | NS/PS/NB | NM/PM/NB |
| | ZE | PM/NB/NB | PS/NM/NB | PS/NM/PB | ZE/NS/ZE | NS/ZE/PB | NM/PS/NB | NM/PM/NB |
| | PS | PS/NB/NB | PS/NM/NB | ZE/NS/PB | NS/NS/ZE | NS/ZE/PB | NM/PS/NB | NM/PS/NB |
| | PM | ZE/NM/NB | ZE/NS/NB | NS/NS/PB | NM/NS/ZE | NM/ZE/PB | NM/PS/NB | NB/PS/NB |
| | PB | ZE/PS/NB | NS/ZE/NB | NS/ZE/PB | NM/ZE/ZE | NM/ZE/PB | NB/PB/NB | NB/PB/NB |

1. $e_2 = x - x_d$

## 6 Stability Analysis

***Theorem 1:*** The control law developed in Equation (39) enables precise trolley positioning at desired coordinates while effectively suppressing swing dynamics of both the hook and payload, i.e.,

$$\lim_{t \to \infty}(x, \dot{x}, \theta_1, \dot{\theta}_1, \theta_2, \dot{\theta}_2) = (x_d, 0, 0, 0, 0, 0). \quad (43)$$

***Proof:*** Construct the Lyapunov equation in the following form:

$$V(t) = \frac{1}{2}\dot{e}^T \dot{e} + E_p \\ = \frac{1}{2}\dot{e}^T \dot{e} + \frac{1}{2}(\frac{g}{l_1} + \frac{m_2 g}{m_1 l_1})\theta_1^2 \\ + K_p \ln \cosh(e_\varepsilon - \frac{\theta_1 + \theta_2}{l_1}) + \frac{3 - \theta_1 \theta_2}{m_1 l_1} m_2 g. \quad (44)$$

Taking the derivative of $V(t)$ and substituting Eq. (39) into it, we obtain:

$$\dot{V}(t) = -K_l \left[\cos\theta_1 \dot{\theta}_1^2 + \cos\theta_2 \dot{\theta}_2^2\right] \\ - \frac{K_d}{m^2 l_1^2} \left(l_1 \dot{e}_\varepsilon - \dot{\theta}_1\right)^2 \le 0. \quad (45)$$

Therefore, $V(t) \le V(0)$, which implies $V(t) \in \mathcal{L}_\infty$, thus

$$e_\varepsilon, \dot{e}_\varepsilon, x, \int_0^t (\sin\theta_1 + \sin\theta_2)\, d\tau, \dot{\theta}_1, \dot{\theta}_2, \dot{x} \in \mathcal{L}_\infty. \quad (46)$$

By combining Eq. (1) and Eq. (39), we have

$$u, \ddot{x}, \ddot{\theta}_1, \ddot{\theta}_2 \in \mathcal{L}_\infty. \quad (47)$$

To apply LaSalle's invariance principle, define the following set:

$$\Gamma = \{(x, \dot{x}, \theta_1, \dot{\theta}_1, \theta_2, \dot{\theta}_2) | \dot{V}(t) = 0\}. \quad (48)$$

Define $\Phi$ as the largest invariant set of $\Gamma$.
In set $\Gamma$, based on Eq. (45), it follows that

$$\dot{\theta}_1 = 0 \quad (49)$$
$$\dot{\theta}_2 = 0. \quad (50)$$

From Eq. (45), (49) and (50),

$$\dot{e}_\varepsilon = \dot{x} - K_l l_1(\sin\theta_1 + \sin\theta_2) = 0. \quad (51)$$

To differentiate Eq. (51) concerning time, we get:

$$\ddot{e}_\varepsilon = \ddot{x} - K_l l_1(\cos\theta_1 \dot{\theta}_1 + \cos\theta_2 \dot{\theta}_2) = 0. \quad (52)$$

From Eq. (49), (50), and (52), we obtain:

$$\ddot{x} = 0 \quad (53)$$
$$\ddot{\theta}_1 = 0 \quad (54)$$
$$\ddot{\theta}_2 = 0. \quad (55)$$

Substituting Eq. (49), (50), (53), (54) and (55) into Eq. (1) results in
$$u = 0. \quad (56)$$
Substituting Eq. (49), (50), (53), (54) and (55) into Eq. (2) and Eq. (3) yields
$$\theta_1 = 0 \quad (57)$$
$$\theta_2 = 0. \quad (58)$$

Substituting Eq. (57), (58) and (54) into Eq. (51) yields
$$\dot{x} = 0. \quad (59)$$

The above derivation has demonstrated that
$$\dot{x}=0, \theta_1=0, \theta_2=0, \dot{\theta}_1=0, \dot{\theta}_2=0. \quad (60)$$

Next, the convergence of $x$ will be discussed.

From Eq. (39), (49), (50), (51), (56), (57) and (58), it follows that
$$e_\varepsilon = 0. \quad (61)$$

It follows from Eq. (8) and Eq. (61) that
$$x - x_d = K_l\, l_1 (\int_0^t \sin\theta_1 d\tau + \int_0^t \sin\theta_2 d\tau). \quad (62)$$

Based on the small angle approximation, $\cos\theta_i = 0$, $\dot{\theta}_2 \sin(\theta_1 - \theta_2) = 0$, according to Eq. (2), it follows that

$$\begin{aligned}&\int_0^t \sin\theta_1 d\tau \\ &= -\frac{1}{(m_1+m_2)gl_1}\int_0^t [(m_1+m_2)l_1\ddot{x}+(m_1+m_2)l_1^2\ddot{\theta}_1+m_2l_1l_2\ddot{\theta}_2]\,d\tau \\ &= -\frac{1}{(m_1+m_2)gl_1}[(m_1+m_2)l_1\dot{x}+(m_1+m_2)l_1^2\dot{\theta}_1+m_2l_1l_2\dot{\theta}_2].\end{aligned} \quad (63)$$

According to Eq. (3), it follows that
$$\begin{aligned}\int_0^t \sin\theta_2 d\tau &= -\frac{1}{m_2gl_2}\int_0^t (m_2l_2\ddot{x}+m_2l_1l_2\ddot{\theta}_1+m_2l_1l_2\ddot{\theta}_2)d\tau \\ &= -\frac{1}{m_2gl_2}(m_2l_2\dot{x}+m_2l_1l_2\dot{\theta}_1+m_2l_1l_2\dot{\theta}_2).\end{aligned} \quad (64)$$

According to Eq. (49), (50) and (59), it follows that
$$\int_0^t \sin\theta_1 d\tau = 0 \quad (65)$$
$$\int_0^t \sin\theta_2 d\tau = 0. \quad (66)$$

From Eq. (62), (65), and (66), we obtain:
$$x = x_d. \quad (67)$$

From Eq. (49), (50), (57), (58), (59), and (67), the following conclusion can be drawn:

The largest invariant set $\Phi$ contains only the equilibrium points. ***Theorem 1*** is porved.

## 7 Simulation Results

This section provides numerical simulation results to validate the performance of the proposed control method.

The system parameters and initial values of the controller parameters in the simulation are as follows:
$$\begin{aligned}&K_{p-initial}=1.5, K_{d-initial}=250, K_{l-initial}=0.01,\\ &x_d=0.7m, m=10kg.\end{aligned} \quad (68)$$

Two cases are considered to conduct the simulations:

**1）Case 1: Under different system parameters.**

To verify the robustness of the control system, two sets of loads(Group 1: $l_1=0.7m$, $m_1=1kg$, $l_2=0.3m$, $m_2=2kg$; Group 1: $l_1=0.7m$, $m_1=1kg$, $l_2=0.4m$, $m_2=1.5kg$) were used, and the results are shown in Fig. 2. It can be observed that, with different payload settings, the trolley can be precisely positioned at the desired location, and the swing angles of the hook and payload converge to zero after oscillating for a period, demonstrating that the controller possesses strong robustness.

**2）Case 2: Comparison with another controller.**

To validate the superior performance of the proposed controller, a comparison with the controller from [17] was conducted. The results are shown in Fig. 3. The results demonstrate that the proposed controller can position the trolley to the desired location more quickly and without oscillations. Compared to the method in [17], the proposed controller is more effective in suppressing the swing angles of the hook and payload within a smaller range. This indicates that the proposed controller offers excellent control performance.

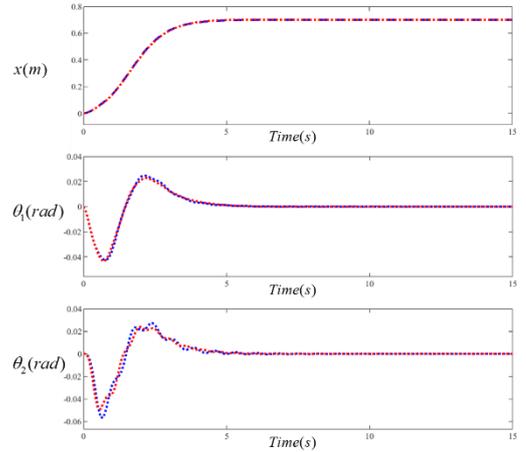

Fig. 2: Simulation Results of Case 1
(red line: $l_1=0.7m$, $m_1=1kg$, $l_2=0.3m$, $m_2=2kg$;
blue line: $l_1=0.7m$, $m_1=1kg$, $l_2=0.4m$, $m_2=1.5kg$).

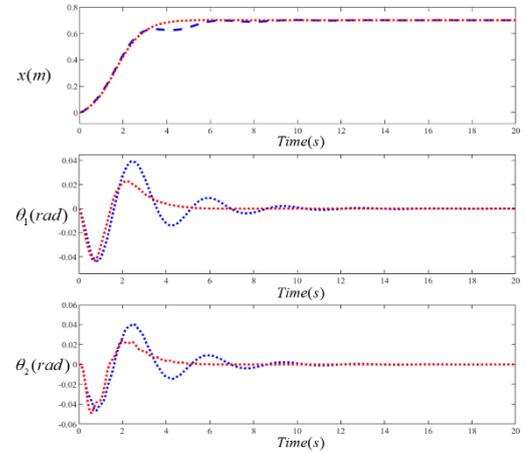

Fig. 3: Fig. 2: Simulation Results of Case 1
(red line: proposed controller;
blue line: controller of [17];
parameters of system: $l_1=0.7m$, $m_1=1kg$, $l_2=0.3m$, $m_2=2kg$).

# 8 Conclusion

This paper addresses the characteristics of underactuated overhead crane systems by coupling the trolley displacement with the swing angles of the hook and load. An enhanced coupling and partially constrained nonlinear controller is proposed. To achieve self-tuning of the controller parameters and improve its performance, a fuzzy logic adaptive strategy is designed to adjust the controller parameters. The detailed design process of the controller and rigorous stability analysis are presented, and the robustness and control performance of the controller are validated through simulations.